# High pCO$_2$ reduces sensitivity to CO$_2$ perturbations on temperate, Earth-like planets throughout most of habitable zone

R.J. Graham

Atmospheric, Oceanic, and Planetary Physics, Clarendon Laboratory, Department of Physics,

University of Oxford, Oxford OX1 3PU, UK

Corresponding Author: R.J. Graham, robert.graham@physics.ox.ac.uk

**Abstract**

The nearly logarithmic radiative impact of $CO_2$ means that planets near the outer edge of the liquid water habitable zone (HZ) require $\sim 10^6$ x more $CO_2$ to maintain temperatures conducive to standing liquid water on the planetary surface than their counterparts near the inner edge. This logarithmic radiative response also means that atmospheric $CO_2$ changes of a given mass will have smaller temperature effects on higher $pCO_2$ planets. Ocean pH is linked to atmospheric $pCO_2$ through seawater carbonate speciation and calcium carbonate dissolution/precipitation, and the response of pH to changes in $pCO_2$ also decreases at higher initial $pCO_2$. Here, we use idealized climate and ocean chemistry models to demonstrate that $CO_2$ perturbations large enough to cause catastrophic changes to surface temperature and ocean pH on low-$pCO_2$ planets in the innermost region of the HZ are likely to have much smaller effects on planets with higher $pCO_2$. Major bouts of extraterrestrial fossil fuel combustion or volcanic $CO_2$ outgassing on high-$pCO_2$ planets in the mid-to-outer HZ should have mild or negligible impacts on surface temperature and ocean pH. Owing to low $pCO_2$, Phanerozoic Earth's surface environment may be unusually volatile compared to similar planets receiving lower instellation.

1. Introduction

Transient perturbations to the carbon cycle that draw down or release large quantities of $CO_2$ are known to have played a central role in Earth's climatic history and the evolution of life. Many of the major and minor extinction events of the Phanerozoic Eon are thought to have been caused in part by global heating and ocean acidification from the $CO_2$ spewed by the eruptions of massive volcanic systems called large igneous provinces (LIPs) [Kidder & Worsley (2010), Sobolev et al. (2011), Bond & Wignall (2014), Bond & Grasby (2017), Ernst & Youbi (2017)]. The most extreme known example of this phenomenon is the Permian-Triassic extinction, which wiped out up to 96% of marine species and likely resulted from the eruption of the Siberian Traps LIP through a large organic carbon deposit [Benton (2018)]. Conversely, cooling periods throughout the Phanerozoic have been driven by tropical

Corresponding Author: R.J. Graham, robert.graham@physics.ox.ac.uk

arc-continent collisions and flood basalt eruptions, each of which can trigger $CO_2$ drawdown by exposing fresh silicates to the surface that are subsequently weathered [Ernst & Youbi (2017), Macdonald et al. (2019)]. Pulses of $CO_2$ consumption may also have served as triggers or preconditions for one or more of the Snowball Earth glaciations that occurred at the ends of the Archean and Proterozoic eons [Donnadieu et al. (2004), Cox et al. (2016), Somelar et al. (2020)]. Furthermore, ongoing anthropogenic $CO_2$ emission is comparable to that of other major carbon release events during the Phanerozoic [Foster et al. (2018)], though much more rapid, and the consequences for human civilization and the biosphere would be severe if the flood of $CO_2$ were allowed to continue until depletion of all fossil fuel resources [Ramirez et al. (2014)].

The above helps make it clear that the Earth's habitability is fundamentally influenced – in some cases, threatened— by large, rapid carbon cycle perturbations [Rothman (2017)]. This motivates a broader analysis of these events to understand their potential importance for the habitability of Earth-like planets in general. Are failed Earths that were heated or frozen by $CO_2$ perturbations until sterilized of complex life common in the galaxy?

On Earth, the sequestration of $CO_2$ by the temperature-dependent chemical conversion of silicate minerals on the continents and seafloor into carbonate minerals is thought to have acted as a first-order negative feedback on climate over geologic timescales [Walker et al. (1981), Coogan & Gillis (2013), Maher & Chamberlain (2014), Graham & Pierrehumbert (2020)]. Under a cooling trend, silicate weathering slows down, allowing $CO_2$ from volcanic outgassing to build to higher levels and oppose the cooling, and the opposite happens in response to warming. This stabilizing feedback modulates climate on $10^5$ - $10^6$ year time intervals [Colbourn et al. (2015)]. Control of $CO_2$ partial pressure via the silicate weathering feedback is likely a primary reason the planet has remained habitable for the past 4 billion years (Ga) [Walker et al. (1981)], despite an initial solar luminosity of about 70% of its modern value and a (1-0.7)/0.7 ≈ 40% increase in luminosity over the same period [Sagan & Mullen (1972), Charnay et al. (2020)]. When the sun was dimmer, $pCO_2$ was higher because

Corresponding Author: R.J. Graham, robert.graham@physics.ox.ac.uk

of the feedback, and the increase in solar brightness has been slow enough for weathering to draw down $CO_2$ and maintain climate in a quasi-steady-state with respect to evolving luminosity.

A working hypothesis that underlies the concept of the classical habitable zone (HZ) (the circumstellar region where $N_2$-$CO_2$-$H_2O$ atmospheres can provide enough warming to maintain stable surface liquid water) is that at least some Earth-like exoplanets within the HZ of their stars can support a silicate weathering feedback like Earth's [Kasting et al. (1988), Kasting et al. (1993), Kopparapu et al. (2013)]. This would imply that this subpopulation of rocky planets will display a trend in $pCO_2$ vs. incoming stellar radiation (``instellation''), with planets that receive less instellation within the HZ generally building up thicker $pCO_2$ atmospheres to compensate for the cooling at lower luminosity [Bean et al. (2017)]. This would allow for the maintenance of liquid water on the surfaces of Earth-like planets for longer periods of time and over a much wider range of conditions than would otherwise be likely.

The radiative forcing response to $CO_2$ is approximately logarithmic, meaning a progressively larger $CO_2$ change is required in order to produce a given amount of forcing as the background $CO_2$ level increases [Pierrehumbert (2010), Huang & Bani Shahabadi (2014)] (the deviation from logarithmic behavior at high $pCO_2$ due to self-broadening does not materially affect this point, see e.g. Fig. 1). This means that the $pCO_2$ required to maintain surface temperatures above freezing rises rapidly as instellation falls, going from parts per million of a bar in the inner portion of the HZ to bars or tens of bars in the mid-to-outer reaches. Given this huge gradient in $pCO_2$ for $N_2$-$CO_2$-$H_2O$ atmospheres across the HZ, changes in $CO_2$ that might cause catastrophe in the inner-HZ should have much smaller impacts in the outer-HZ because of the logarithmic forcing effect. For example, a LIP that suddenly increased the atmospheric $CO_2$ concentration on an inner-HZ planet from a few hundred parts per million by volume (ppmv) to a few thousand ppmv would likely cause severe warming and (in the presence of life) a mass extinction, but adding the same mass of $CO_2$ to an atmosphere in the outer-HZ that already had a $pCO_2$ of a few bars would result in a negligible change in surface temperature.

Corresponding Author: R.J. Graham, robert.graham@physics.ox.ac.uk

In this article, we use an idealized climate model to quantify the variation in temperature impact of transient carbon perturbations to the atmospheres of Earth-like planets with different $pCO_2$ values at a range of instellations throughout the HZ. We also apply a basic model of carbonate equilibrium and calcium carbonate buffering in the sea to estimate the extent of ocean acidification from these perturbations, another of the important killing mechanisms in mass extinctions caused by LIPs on Earth [Benton (2018)]. In Section 2, we introduce the models used in the article. In Section 3, we present the results of the calculations, demonstrating extreme climate resilience to carbon cycle perturbations for temperate planets in the mid-to-outer reaches of the HZ. In section 4, we discuss caveats and implications of these results, and in section 5 we summarize and conclude the study.

## 2. Modeling & Methods

### 2.1. Energy balance climate model

To investigate the climate sensitivity of Earth-like planetary climate to carbon cycle perturbations at varying instellations, we examine the changes to global-mean surface temperature from atmospheric $CO_2$ increases equivalent to 5000 gigatonnes of carbon (GtC) and 50,000 GtC. A perturbation of 5000 GtC is on the order of the size of perturbation that caused the Paleocene-Eocene Thermal Maximum [McInerney & Wing (2011)] and is approximately equal to estimates of total global hydrocarbon resources (which includes both accessible and currently-inaccessible reservoirs of fossil fuels) [Rogner (1997), Ramirez et al. (2014)]. A perturbation of 50,000 GtC is of the same order of magnitude as the carbon release caused by the Siberian Traps during the Permian-Triassic mass extinction [Svensen et al. (2009)]. We use an ``inverse climate modeling'' approach, as described in e.g. [Kasting (1991)], to calculate the initial pre-perturbation climate in each case: the surface temperature and $pCO_2$ are first assumed, which allows calculation of the albedo and outgoing longwave radiation (OLR), which can together be used to calculate the effective stellar flux ($S_{eff}$; the ratio of incoming stellar flux, $S$, on a given planet to modern Earth's insolation of $S_{earth}$=1368 W m$^{-2}$)

Corresponding Author: R.J. Graham, robert.graham@physics.ox.ac.uk

required to sustain the assumed global-mean temperature-$pCO_2$ combination under the assumption of global-mean energy balance:

$$S_{\text{eff}} = \frac{S}{S_{\text{earth}}} = \frac{4 \times OLR(T, pCO_2)}{S_{\text{earth}} \times (1 - a(T, pCO_2))} \qquad (1)$$

where $a(T, pCO_2)$ is the planetary albedo and all other symbols have been introduced above. This highly idealized, global-mean approach is justified for this study because the climate effect we demonstrate results from the basic, qualitative behavior of atmospheric $CO_2$.

To calculate OLR as a function of $pCO_2$ and global-mean surface temperature, we apply a piecewise 6[th]-degree polynomial function developed by [Kadoya & Tajika (2019)] to approximate simulations of an $N_2$-$CO_2$-$H_2O$ atmosphere by the 1D radiative-convective column model used in [Haqq-Misra et al. (2016)]. The fit can accurately (with error <=3.3 W m$^{-2}$) reproduce the column model's OLR output for $10^{-5}$ bar < $pCO_2$ < 10 bar and 150 K < $T$ < 350 K. For all $pCO_2$ and $T$, the atmosphere has a constant $pN_2$ = 1 bar and is saturated with $H_2O$. OLR output from this model is plotted as a function of temperature over the range 273 K < $T$ < 340 K and $pCO_2$ over the range $10^{-4}$ bar < $pCO_2$ < 10 bar in Fig. 1. In matrix form, the equation is:

$$OLR(T, pCO_2) = I_0 + \mathbf{TBY}^t \qquad (2)$$

$$\mathbf{T} = (1 \; \xi \; \xi^2 \; \xi^3 \; \xi^4 \; \xi^5 \; \xi^6) \qquad (3)$$

$$\mathbf{Y} = (1 \; \upsilon \; \upsilon^2 \; \upsilon^3 \; \upsilon^4) \qquad (4)$$

where $T$ is the surface temperature in Kelvin, $pCO_2$ is the partial pressure of $CO_2$ in bars, and $\xi$ = 0.01x($T$-250). For $pCO_2$ < 1 bar,

$$p = 0.2 \times \log_{10}(pCO_2) \qquad (5)$$

$$\mathbf{B} = \begin{bmatrix} 87.8373 & -311.289 & -504.408 & -422.929 & -134.611 \\ 54.9102 & -677.741 & -1440.63 & -1467.04 & -543.371 \\ 24.7875 & 31.3614 & -364.617 & -747.352 & -395.401 \\ 75.8917 & 816.426 & 1565.03 & 1453.73 & 476.475 \\ 43.0076 & 339.957 & 996.723 & 1361.41 & 612.967 \\ -31.4994 & -261.362 & -395.106 & -261.600 & -36.6589 \\ -28.8846 & -174.942 & -378.436 & -445.878 & -178.948 \end{bmatrix} \qquad (6)$$

Corresponding Author: R.J. Graham, robert.graham@physics.ox.ac.uk

and when pCO$_2$ exceeds 1 bar, $p$ and $\mathbf{B}$ are instead:

$$p = \log_{10}(\text{pCO}_2) \tag{7}$$

$$\mathbf{B} = \begin{bmatrix} 87.8373 & -52.1056 & 35.2800 & -1.64935 & -3.43858 \\ 54.9102 & -49.6404 & -93.8576 & 130.671 & -41.1725 \\ 24.7875 & 94.7348 & -252.996 & 171.685 & -34.7665 \\ 75.8917 & -180.679 & 385.989 & -344.020 & 101.455 \\ 43.0076 & -327.589 & 523.212 & -351.086 & 81.0478 \\ -31.4994 & 235.321 & -462.453 & 346.483 & -90.0657 \\ -28.8846 & 284.233 & -469.600 & 311.854 & -72.4874 \end{bmatrix} \tag{8}$$

with symbols retaining their previous meanings. OLR is plotted as a function of temperature over the range 273 K < $T$ < 340 K and pCO$_2$ over the range 10$^{-4}$ bar < pCO$_2$ < 10 bar in Fig. 1.

We calculate global-mean planetary albedo using another polynomial fit to radiative-convective model output for a planet orbiting a Sun-like star with T$_{eff}$=5800 K, sourced from the same model the OLR equation is fit to, though in this case the albedo parameterization comes from [Haqq-Misra et al. (2016)]:

$$\begin{aligned}\alpha(t, a_s, \mu, \phi) = {} & A\mu^3 + B\mu^2 a_s + C\mu^2 t + D\mu^2\phi + E\mu^2 + F\mu a_s^2 + G\mu a_s t + H\mu a_s \phi \\ & + I\mu a_s + J\mu t^2 + K\mu t\phi + L\mu t + M\mu\phi^2 + N\mu\phi + O\mu + P a_s^3 \\ & + Q a_s^2 t + R a_s^2 \phi + S a_s^2 + T_0 a_s t^2 + U a_s t\phi + V a_s t + W a_s \phi^2 \\ & + X a_s \phi + Y a_s + Z t^3 + AA t^2 \phi + AB t^2 + AC t\phi^2 + AD t\phi + AE t \\ & + AF\phi^3 + AG\phi^2 + AH\phi + AG \end{aligned} \tag{9}$$

where $t$ = log$_{10}$($T$), $a_s$ is the surface albedo (taken to be 0.3 by default), $\mu$ is the cosine of the stellar zenith angle (taken to be 2/3, the instellation-weighted average, following [Cronin (2014)]), and $\phi$ = log$_{10}$(pCO$_2$). The coefficients $A$ through $AG$ in the equation can be downloaded as a [tar.gz](tar.gz) file made available by Haqq-Misra et al. The file includes coefficients for stars with stellar effective temperatures of 3400 K, 4600 K, 5800 K (which we use), and 7200 K, since different stellar spectra lead to different planetary albedos. The coefficient values are also listed in Table 1.

Corresponding Author: R.J. Graham, robert.graham@physics.ox.ac.uk

Table 1: The coefficients in Eq. (9) for albedo

| Label | Coefficient value | Label | Coefficient value |
|---|---|---|---|
| A | -4.41391620e-01 | S | 6.11699085e-01 |
| B | -2.60017516e-01 | $T_0$ | -2.33213410e+00 |
| C | 1.08110772e+00 | U | 2.56011431e-01 |
| D | -3.93863286e-02 | V | 1.05912148e+01 |
| E | -1.46383456e+00 | W | -1.85772689e-02 |
| F | 9.91383779e-02 | X | -7.55796861e-01 |
| G | -1.45914724e+00 | Y | -1.16485004e+01 |
| H | -2.72769393e-02 | Z | 2.74062492e+01 |
| I | 3.99933641e+00 | AA | 5.46044241e-01 |
| J | 1.07231336e+00 | AB | -2.05761674e+02 |
| K | -1.04302521e-02 | AC | 5.57943359e-02 |
| L | -6.10296439e+00 | AD | -2.49880330e+00 |
| M | 2.69255204e-03 | AE | 5.14448995e+02 |
| N | 9.50143253e-02 | AF | 2.43702089e-03 |
| O | 7.37864216e+00 | AG | -1.09384841e-01 |
| P | 1.28580729e-01 | AH | 2.92643187e+00 |
| Q | -3.07800301e-01 | AG | -4.27802455e+02 |
| R | 2.27715595e-02 | | |

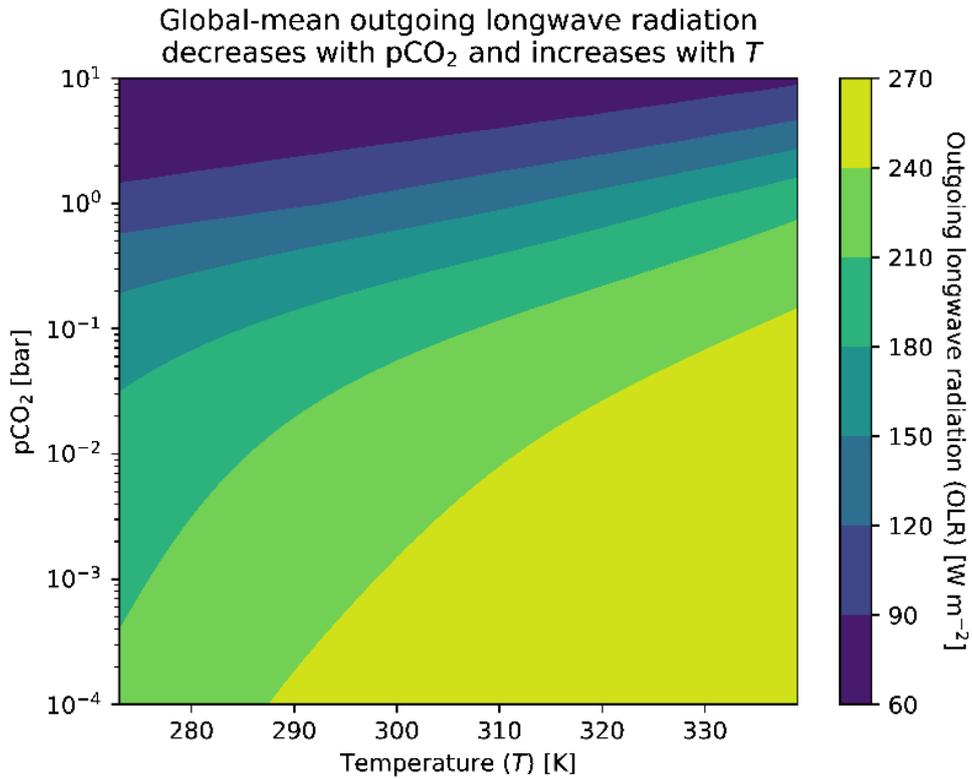

Fig. 1: OLR as a function of pCO$_2$ and global-mean surface temperature $T$.

Corresponding Author: R.J. Graham, robert.graham@physics.ox.ac.uk

Increasing pCO$_2$ elevates albedo due to CO$_2$'s strong Rayleigh scattering cross-section, and increasing temperature tends to reduce albedo by increasing the concentration of H$_2$O, which absorbs longwave and some shortwave radiation, reducing the effectiveness of scattering. Albedo is plotted as a function of temperature over global-mean surface temperature range 273 K < $T$ < 340 K and pCO$_2$ range $10^{-4}$ bar < pCO$_2$ < 10 bar in Fig. 2. The slight reduction in albedo visible in Fig. 2 that occurs with a pCO$_2$ increase from $10^{-4}$ bar to $10^{-3}$ bar at a temperature of around 325 K is a spurious result of an imperfect polynomial fit, but the simulations in this study never access that region of the parameter space, so it does not impact the results presented here.

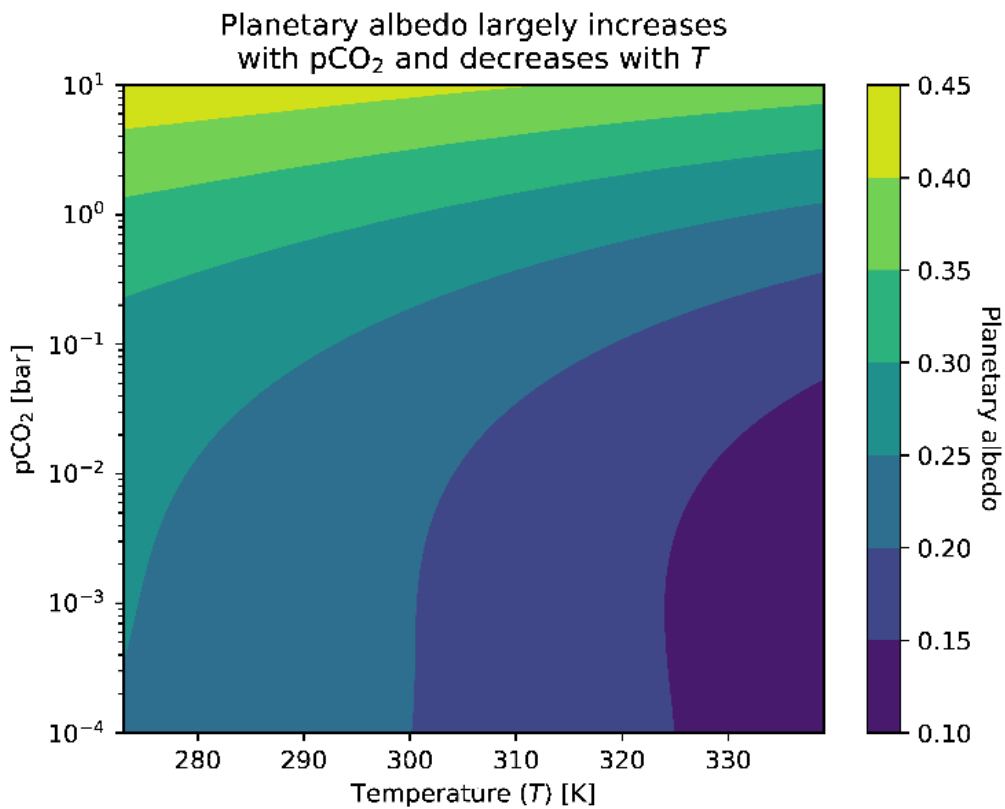

**Fig. 2:** Planetary albedo as a function of pCO$_2$ and global-mean surface temperature $T$.

With boundary conditions characteristic of pre-industrial Earth, e.g. an instellation of $S$ = 1368 W m$^{-2}$ and pCO$_2$ = 280 x 10$^{-6}$ bar, the model produces $T$ = 309 K, significantly hotter than Earth's pre-industrial temperature of ~285 K [Walker et al. (1981)]. This is at least partly because the model

Corresponding Author: R.J. Graham, robert.graham@physics.ox.ac.uk

atmosphere is assumed to be saturated with water vapor, whereas Earth's atmosphere is held at global-mean relative humidities less than 1 due to regions of dry, subsiding air, allowing for more efficient radiation to space and thus cooler surface temperatures e.g. [Pierrehumbert (1995)]. In addition, cloud feedbacks that can have important impacts on OLR and albedo are not included in this model. Cloud effects can alter long term climate and the relationship between $pCO_2$ and temperature at a given instellation on Earth-like planets e.g. [Goldblatt et al. (2021)], but they are intrinsically 3-dimensional phenomena, so they require 3-D general circulation models to simulate. To ensure the simplifications noted above are minor for the purposes of this study, next we evaluate the sensitivity of the climate model to ensure it reproduces the relevant behavior of more sophisticated models.

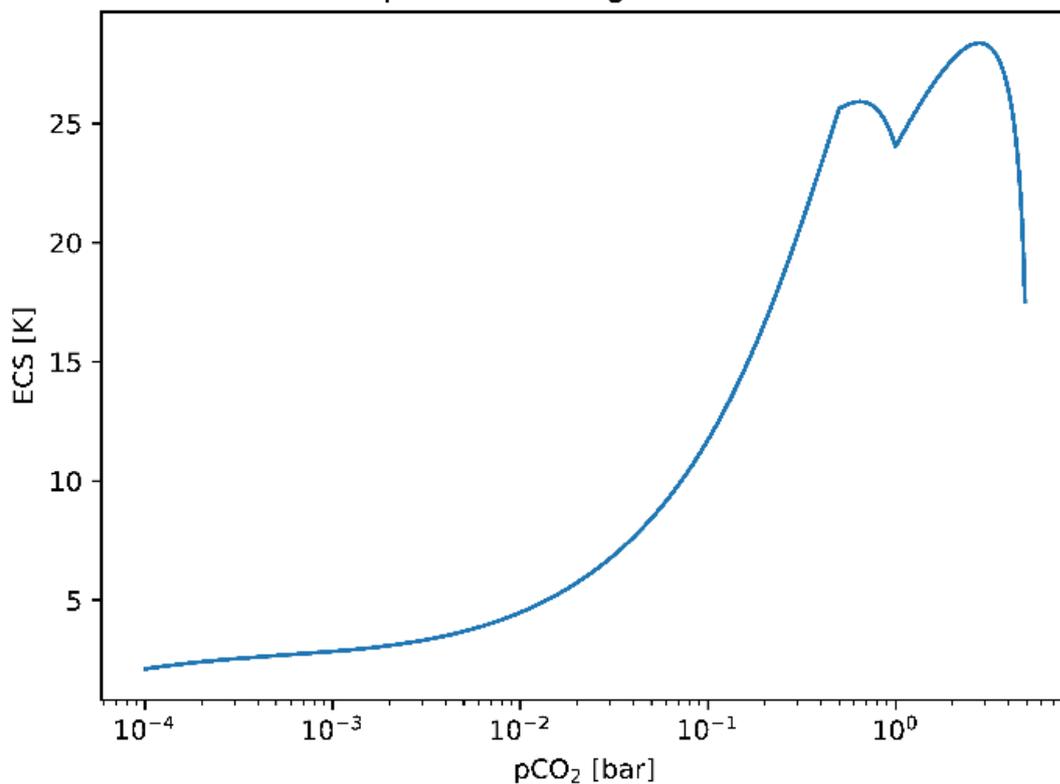

Fig. 3: ECS as a function of $pCO_2$ for planets with initial $T$=273.15 K.

Corresponding Author: R.J. Graham, robert.graham@physics.ox.ac.uk

The steady-state global-mean temperature increase produced by a doubling of $CO_2$ is known as the equilibrium climate sensitivity (ECS) and is a common metric in studies of Earth climate. In more sophisticated climate models than the model used in this paper, the value of ECS is dependent on many processes in addition to the radiative effects of $CO_2$ (and $H_2O$), particularly the dynamics of low-lying, high-albedo clouds [Wolf et al. (2018), Schneider et al. (2019)]. Based on climate models of varying complexity, ECS on Earth may vary by an order of magnitude depending on background conditions, from ~2 K to ~20 K, with a general trend toward higher sensitivity as $pCO_2$ increases and a pronounced maximum at warm temperatures and intermediate $pCO_2$ that arises from low-lying cloud dissipation [Wolf et al. (2018)] and clear-sky radiative feedbacks [Seeley & Jeevanjee (2020)]. To ensure that the model in this paper produces reasonable climatic responses to $CO_2$ changes, we examine the temperature perturbations from $pCO_2$ doublings with initial $pCO_2$ values ranging from $10^{-4}$ bar to 5 bars (beyond which a $CO_2$ doubling takes the model outside of its range of validity in $pCO_2$ space), with each simulation initialized at $T$=273.15 K (see Fig. 3). The range of ECS values produced by our climate model is broadly consistent with the range of values in the literature (e.g. the ECS throughout Earth history is estimated to have varied between 2.6 K and 21.6 K by [Wolf et al. (2018)]), though there is no maximum at intermediate $pCO_2$ since the calculations are all carried out with low initial temperatures. The model does reproduce the intermediate-$pCO_2$ ECS maximum when initialized at $T \geq$ ~300 K, (not shown). At $pCO_2 = 10^{-4}$ bar, ECS ~ 2 K, which increases to a maximum of around 27 K near 2 bars. These numbers vary by a few degrees in either direction when the model is initialized from different temperatures, but the differences are small relative to the effects demonstrated in this study. The kink in the ECS that occurs at 1 bar is due to the discontinuity at $pCO_2$= 1 bar in the OLR polynomial fit from [Kadoya & Tajika (2019)] used in this climate model. $CO_2$'s self-broadening increases ECS as background $pCO_2$ goes up, meaning the model's radiative behavior is not perfectly logarithmic (ECS would be constant in that case), but ECS varies by approximately an order of magnitude across five orders of magnitude of $pCO_2$, confirming the usefulness of the logarithmic approximation in discussing the model's behavior. ECS begins to decrease rapidly beyond

Corresponding Author: R.J. Graham, robert.graham@physics.ox.ac.uk

about 2 bars because of the increasing relative impact of $CO_2$'s Rayleigh scattering on the energy balance of the planet. If the curve were extended until ECS intercepted the x-axis, that point would define the ``maximum greenhouse limit,'' beyond which the addition of $CO_2$ to the atmosphere would cool the planet e.g. [Kopparapu et al. (2013)].

We first determine the $S_{eff}$ necessary at a given $pCO_2$ to hold a planet's global-mean temperature at $H_2O$'s freezing point, $T$=273.15 K, which serves as an approximate minimum temperature & $pCO_2$ for surface habitability at a given instellation. This is a conservative choice for addressing the question of the impact of carbon cycle perturbations, as it minimizes initial atmospheric $CO_2$, maximizing the impact of a given $pCO_2$ increase. Then, given the initial $CO_2$, we calculate the $pCO_2$ increase from the addition of a mass of $CO_2$ $m_{CO2}$:

$$\Delta pCO_2 = \frac{\mu_{atm}}{\mu_{CO2}} \frac{m_{CO2} g}{4\pi R_{earth}^2} \qquad (10)$$

where $\Delta pCO_2$ is the change in $CO_2$ partial pressure, $\mu_{atm}$, which varies with relative proportions of $CO_2$ and $N_2$ ($\mu_{N2}$=28 g mol$^{-1}$), is the molar mass for the atmosphere as a whole, $\mu_{CO2}$=44 g mol$^{-1}$ is the molar mass for $CO_2$, $g$=9.81 m s$^{-2}$ is the acceleration due to gravity on Earth (we assume planets are the same size as Earth in this article), and $R_{earth}$=6.371x10$^6$ m is the radius of the Earth. Finally, to calculate the temperature change $\Delta T$ from a perturbation, we simply add the $pCO_2$ increase to the initial $pCO_2$ and calculate the new temperature necessary to maintain energy balance at the original instellation ($S$).

## 2.2. Oceanic carbonate system and pH

To estimate the impact of the calculated changes to $pCO_2$ and $T$ on seawater pH, we assume the oceanic carbonate system is in equilibrium with atmospheric $CO_2$ and with calcium carbonate that buffers changes to pH by dissolving or precipitating. The following equations and constants are drawn from [Pierrehumbert (2010)], but with the concentrations converted from units of [mol mol$^{-1}$] to units of [mol L$^{-1}$] through multiplication by a factor of 55.56 mol L$^{-1}$, the number of moles of liquid water

Corresponding Author: R.J. Graham, robert.graham@physics.ox.ac.uk

per liter. All constants are set to values appropriate for an ocean depth of 3 km, with salinity approximately equal to 35 grams of salt per kilogram of water. The solubility equilibrium equation for CaCO$_3$'s dissociation into Ca$^{2+}$ and CO$_3^{2-}$ is:

$$K_{\text{sp}} = [\text{Ca}^{2+}][\text{CO}_3^{2-}] \quad (11)$$

where $K_{\text{sp}}$= 7.90 x 10$^{-7}$ mol$^2$ L$^{-2}$ is CaCO$_3$'s solubility product and each bracketed chemical species represents the concentration of that chemical, written as a molarity. The equilibrium equation for the reaction H$_2$O + CO$_2$ (aq) ↔ H$^+$(aq) + HCO$_3^-$ (aq) can be written:

$$K_1(T) = \frac{[\text{H}^+][\text{HCO}_3^-]}{[\text{CO}_2]} = K_{1,\text{ref}} \times \exp\left(-c_1\left(\frac{1}{T} - \frac{1}{298}\right)\right) \quad (12)$$

where $K_1(T)$ is the temperature-dependent equilibrium constant, $K_{1,\text{ref}}$ = 1.32 x 10$^{-6}$ mol L$^{-1}$ is the value of $K_1$ at 298 K, and $c_1$ = 1218 K is the temperature dependence of the reaction. The equilibrium equation for the reaction HCO$_3^-$ (aq) ↔ H$^+$ (aq) + CO$_3^{2-}$ (aq) can be written:

$$K_2(T) = \frac{[\text{H}^+][\text{CO}_3^{2-}]}{[\text{HCO}_3^-]} = K_{2,\text{ref}} \times \exp\left(-c_2\left(\frac{1}{T} - \frac{1}{298}\right)\right) \quad (13)$$

where the symbols have the same meanings as before (for their respective reaction), $K_{2,\text{ref}}$=8.95 x 10$^{-10}$ mol L$^{-1}$, and $c_2$ = 2407 K. Henry's law for CO$_2$ is written:

$$\text{pCO}_2 = k_\text{H}[\text{CO}_2] \quad (14)$$

where $k_\text{H}$= 29 bar L mol$^{-1}$ is the Henry's law constant for CO$_2$, which relates its partial pressure to its concentration in water at equilibrium. Lastly, under the simplifying assumption that Ca$^{2+}$ is the only source of alkalinity in the ocean, charge balance can be written:

$$2[\text{Ca}^{2+}] = [\text{HCO}_3^-] + 2[\text{CO}_3^{2-}] + \frac{10^{-14}}{[\text{H}^+]} - [\text{H}^+] \quad (15)$$

where $\frac{10^{-14}}{[\text{H}^+]}$ is equal to [OH$^-$].

Corresponding Author: R.J. Graham, robert.graham@physics.ox.ac.uk

Combined and rearranged, the previous five equations produce a convenient fourth degree polynomial relating [H$^+$] to pCO$_2$ and $T$:

$$2K_{sp}[H^+]^4 + \frac{K_1(T)K_2(T)}{k_H}pCO_2[H^+]^3 \\ - \left(\frac{K_1(T)^2 K_2(T)}{k_H^2}pCO_2^2 + \frac{K_1(T)K_2(T)}{k_H}10^{-14}pCO_2\right)[H^+] \\ - 2\frac{K_1(T)^2 K_2(T)^2}{k_H^2}pCO_2^2 = 0 \quad (16)$$

which can be solved for [H$^+$] (and therefore pH, equal to -log$_{10}$[H$^+$]) numerically using the pCO$_2$ and $T$ output from the climate model. It is worth noting that we are implicitly assuming a higher mass of CO$_2$ enters the atmosphere-ocean system than $m_{CO2}$ alone by including ocean carbonate chemistry changes in response to atmospheric CO$_2$ because some carbon must be diverted into the ocean for changes to pH to occur. This effect becomes smaller as the initial pCO$_2$ inventory increases farther from the star, since increasing CO$_2$ causes a larger and larger fraction of the total carbon in the atmosphere-ocean system to be partitioned into the atmosphere [Pierrehumbert (2010)]. With preindustrial Earth-like conditions of pCO$_2$ = 280 x 10$^{-6}$ bar and $T$ = 285 K, the model produces a pH of 8.32, which is reasonably close to Earth's preindustrial pH of ~8.25 [Jacobson (2005)].

3. Results

In Section 3.1, we describe the temperature impact of increases in atmospheric CO$_2$ of 5000 GtC and 50,000 GtC on planets orbiting G-type stars throughout the HZ. In Section 3.2, we describe the change to oceanic pH from the same CO$_2$ increases.

3.1. Temperature change from CO$_2$ perturbations

For both the 5000 GtC perturbation and the 50,000 GtC perturbation, temperature changes go from very large in the inner portion of the HZ to negligibly small in the middle and outer reaches (see middle row of plots in Fig. 4). An initial pCO$_2$ inventory of 10$^{-4}$ bar is the lowest used in this study, equivalent to a concentration of 100 parts per million by volume (ppmv) of CO$_2$ in a 1 bar N$_2$

Corresponding Author: R.J. Graham, robert.graham@physics.ox.ac.uk

atmosphere, and this pCO$_2$ produces an initial $T$ = 273.15 K at $S_{eff}$=0.846. In this case the planet experiences a pCO$_2$ increase of more than an order of magnitude from the 5000 GtC injection, to 2.3 x 10$^{-3}$ bar (2300 ppmv; see top-left panel of Fig. 4). This is accompanied by a temperature increase of 15 K (see mid-left panel of Fig. 4). In the 50,000 GtC case, the final pCO$_2$=2.3x10$^{-2}$ bar (23,000 ppmv; top-right panel of Fig. 4), with a temperature increase of 36 K (mid-right panel of Fig. 4). For comparison, on Earth, LIP eruptions have been associated with temperature increases of up to 15 K [Ernst & Youbi (2017)].

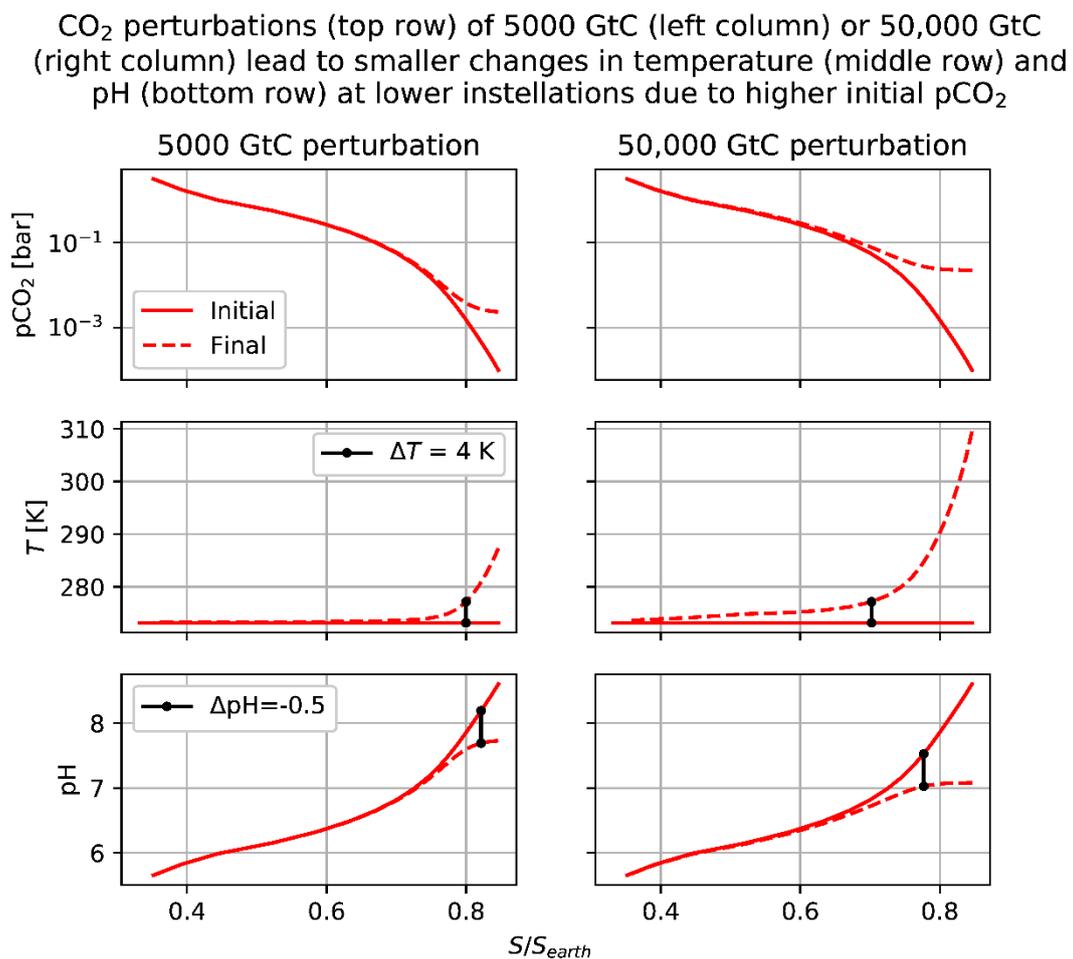

**Fig. 4:** Comparison of temperature-change and pH-change for CO$_2$ perturbations of 5000 GtC (left column) and 50,000 GtC (right column). The x-axis in each case is the instellation relative to Earth's, $S_{eff}$=$S$/$S_{earth}$. The top row shows the pre-perturbation (solid curve) and post-perturbation pCO$_2$ (dashed curve) for planets that are initialized at $T$=273.15 K. The middle row shows the temperature change from the CO$_2$ perturbations, with the solid and dashed curves representing the pre- and post-perturbation temperature values and the black line segment marking the point where the temperature change falls below 4 K. The bottom row shows the oceanic pH change from the CO$_2$

Corresponding Author: R.J. Graham, robert.graham@physics.ox.ac.uk

perturbations, with the solid and dashed curves representing the pre- and post-perturbation pH values and the black line segment marking where the magnitude of acidification falls below 0.5 pH units.

At the outermost orbit evaluated in this study, $S_{eff}$ is equal to 0.352 and a $pCO_2$ of 3.162 bars is necessary to initialize at $T$=273.15 K. For that configuration, a 5000 GtC injection elevates $pCO_2$ to 3.166 bars and increases the temperature by 0.04 K, a change approximately 3 orders of magnitude smaller than the change at the innermost orbit. A 50,000 GtC injection causes a temperature increase of 0.4 K and increases $pCO_2$ to 3.19 bars.

By choosing a semi-arbitrary level of $\Delta T$ as the threshold between a ``large'' and a ``small'' climate change, it's possible to define orbital boundaries beyond which planets inhabit ``stability zones'' where they experience small changes in response to $CO_2$ increases of a given size. Here, we use $\Delta T$ = 4 K as the threshold value separating large from small, as 4 K of heating by 2100 is a worst-case scenario for anthropogenic climate change which would lead to widespread environmental disaster [New et al. (2011)]. For the 5000 GtC $CO_2$ injection, $\Delta T \leq 4$ K at $S_{eff} \leq 0.80$, where the initial $pCO_2$ was $1.5 \times 10^{-3}$ bar. For planets orbiting a star with the Sun's properties where $S_{eff}$=1 at an orbital radius of 1 astronomical unit (au), this translates to climate stability against 5000 GtC perturbations beyond an orbital distance of $d = \sqrt{1/0.80} = 1.12$ au (see boundary between the orange and yellow rings in Fig. 5). For the 50,000 GtC perturbation, climate stability occurs at $S_{eff} \leq 0.70$ (with an initial $pCO_2$ of $5.4 \times 10^{-2}$ bar), translating to an inner orbital distance of $d = 1.20$ au for the 50,000 GtC stability zone (see boundary between yellow and green rings in top panel of Fig. 5).

The inner edge of the HZ for rapidly-rotating planets is usually defined as the instellation where climates are forced into either a "moist greenhouse" state [Kasting et al. (1984)] or a "runaway greenhouse" state [Ingersoll (1969)], given some assumptions about a planet's inventory and surface distribution of water. The moist greenhouse state occurs when a planet's stratosphere becomes wet, triggering efficient photodissociation and hydrogen loss that in time depletes a planet of its oceans [Kasting et al. (1984)]. This can occur at a variety of instellations depending on background $pCO_2$ and

Corresponding Author: R.J. Graham, robert.graham@physics.ox.ac.uk

the atmospheric history of the planet, with values as low as $S_{eff}$=1.03 to 1.05 recovered in GCM simulations of a rapidly rotating aquaplanet around a Sun-like star [Popp et al. (2016)]. The runaway greenhouse occurs when a planet's absorbed instellation exceeds the maximum OLR the planet can radiate in the presence of a condensed reservoir of greenhouse gas in vapor equilibrium with the atmosphere e.g. [Goldblatt & Watson (2012)], leading to uncontrollable warming until the condensable reservoir has evaporated. The runaway greenhouse limit is about $S$ = 375 W m$^{-2}$ for the Sun [Leconte et al. (2013)], which translates to $S_{eff}$=1.1 and $d = 0.95$ au. As the runaway greenhouse is generally thought to occur closer to the star than the moist greenhouse, we will take the runaway limit as the inner edge of the HZ for our calculations. The outer edge of the HZ, where added $CO_2$ cools the planet below freezing instead of warming it, is at approximately $d = 1.67$ au in our solar system [Kopparapu et al. (2013)].

With the above choices of inner and outer HZ boundaries, the innermost ring of the HZ where planets are susceptible to both 5000 GtC and 50,000 GtC perturbations has a radial width of 0.17 au, while planets are safe from 5000 GtC perturbations over a 0.55 au range of radii, and planets are safe from both 5000 and 50,000 GtC perturbations across 0.47 au. An alternative visualization of the climate stability is provided in the bottom panels of Fig. 5, which show the temperature responses to the 5000 GtC (bottom-left panel) and 50,000 GtC (bottom-right panel) perturbations as functions of orbital radius. In the 5000 GtC case, the temperature response is less than 1 K throughout most of the HZ, while for the 50,000 GtC perturbation the response is still modest but stays above 1 K until near the outer edge of the HZ. So, throughout most of the classical HZ, temperate planets where $CO_2$ & $H_2O$ are the dominant greenhouse gases will lie in stability zones where global surface temperature should only be mildly affected by transient carbon cycle perturbations of similar size to those that have repeatedly caused large, catastrophic swings in Earth's surface temperature during the Phanerozoic Eon, and potentially the Neoproterozoic. It is worth noting that if we had assumed a higher initial temperature and a correspondingly higher initial p$CO_2$ at each luminosity, then the inner boundaries for the stable zones would be even closer to the star, so the choice of initializing at

Corresponding Author: R.J. Graham, robert.graham@physics.ox.ac.uk

$T$=273.15 K is conservative. In any case, this calculation is meant to be illustrative; the same qualitative pattern emerges regardless of the choice of $T$, as long as $CO_2$ is assumed to be the dominant greenhouse gas offsetting the reduction in solar luminosity with orbital distance.

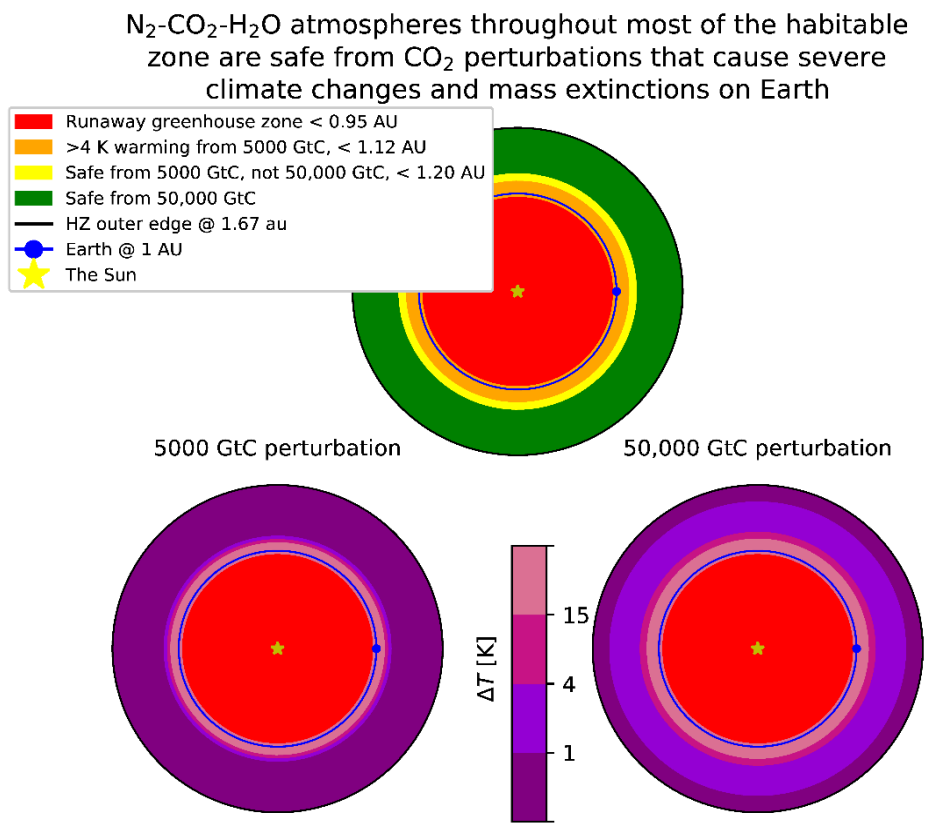

**Fig. 5:** Illustrations of the climate stability as a function of orbital radius within the habitable zone in a solar system with a star identical to the Sun. The yellow star in the center of each panel represents a Sun-like G-type star. The red zone in each panel is the region where Earth-like planets are likely to be in the uninhabitable runaway greenhouse state, extending to 0.95 au. The circular blue dot and ring in each panel represent the Earth's orbit, at a radius of 1 au. In the top panel, the orange zone is within the habitable zone but outside of either stability zone, meaning Earth-like planets in this region would have the potential to warm by more than 4 K under a 5000 GtC $CO_2$ perturbation, based on the temperature response of simulations initialized at $T$=273.15 K. The orange zone extends from the inner edge of the HZ at 0.95 au to 1.12 au. The yellow zone represents the region where planets within the HZ would experience less than 4 K of warming from a 5000 GtC perturbation but more than 4 K of warming from a 50,000 GtC perturbation, and it extends from 1.12 au to 1.20 au. The green zone represents the region where planets within the HZ would experience less than 4 K of warming from either a 5000 GtC perturbation or a 50,000 GtC perturbation, extending from 1.20 au to 1.67 au. The black region is beyond the outer edge of the HZ, beginning at 1.67 au. The bottom panels show the temperature change as a function of orbital radius for the 5000 GtC perturbation (left) and the 50,000 GtC perturbation (right).

Corresponding Author: R.J. Graham, robert.graham@physics.ox.ac.uk

### 3.2. pH change from CO₂ perturbations

Similar to the temperature perturbation results, the change in pH from 5000 and 50,000 GtC atmospheric $CO_2$ mass injections is large only in the inner region of the HZ (see bottom row in Fig. 4). Initial pH is smaller at lower $S_{eff}$ because of higher initial $pCO_2$ values. As $S_{eff}$ decreases, so does the reduction in pH from a given $CO_2$ increase, going from a change of nearly 2 pH units (e.g. a 2 order-of-magnitude increase in [$H^+$]) under the 50,000 GtC increase at the innermost orbit to a change of less than one hundredth of a pH unit at the outermost orbit evaluated.

Stability zones for pH can be defined in the same way we defined them for temperature. In this case, we will use -0.5 pH units as the threshold between a ``large'' and ``small'' pH change, since that is the estimated level of oceanic pH reduction below the pre-industrial value in the year 2100 for a worst-case scenario of climate change, and represents a level of change that would pose a danger to many extant marine animal taxa [Wittmann & Pörtner (2013)]. With that choice of threshold, under a 5000 GtC perturbation, the boundary between safety and vulnerability is at $S_{eff}$ = 0.82, e.g. 1.10 au from the star when orbiting a host identical to our Sun, with an initial $pCO_2$=4.6 x $10^{-4}$ bar (see black line segment in the bottom-left panel of Fig. 4). At 50,000 GtC, the boundary moves out to $S_{eff}$=0.78, or 1.13 au, with an initial $pCO_2$=4.5 x $10^{-3}$ bar (black line segment in the bottom-right panel of Fig. 4). The pH stability zones for 5000 GtC and 50,000 GtC both lie closer to the star than their temperature-stability analogues, so in the rest of the article, ``stability zone'' will refer to the more conservative temperature-based zones.

### 4. Discussion

#### 4.1. Mechanisms of catastrophe & mass extinction

Hyperthermals and potentially one or more of the snowball events in Earth's past were caused by transient carbon cycle perturbations, and these events strongly influenced the particular evolutionary path taken by the biosphere. Significant warming and/or ocean acidification, often as a result of the eruption of LIPs, is associated with most or all of the major mass extinctions in the Phanerozoic [Bond

Corresponding Author: R.J. Graham, robert.graham@physics.ox.ac.uk

& Grasby (2017), Lindström et al. (2017), Benton (2018), Racki et al. (2018), Schoene et al. (2019), Bond & Grasby (2020)]. The stability of global-mean surface temperature and ocean pH against transient carbon cycle perturbations on planets in the mid-to-outer reaches of the HZ suggests that such catastrophic climate events and mass extinctions may happen much less frequently for this class of Earth-like planet than they have on Earth. However, mass extinctions and climate perturbations can be caused by mechanisms other than temperature- and pH-change from $CO_2$ perturbations. It is worth examining the mechanisms underlying the catastrophes in Earth's past for context to understand the implications of the results in this paper.

A relevant category of catastrophic environmental processes is the set that can or must be triggered by an initial pulse of warming. Processes of this type have been important in past mass extinction events, and the resistance to $CO_2$-based temperature change on mid-outer HZ planets should make them less likely. Ocean anoxia is one such process. Ocean anoxia can be caused by increased continental weathering—a result of elevated temperature and/or $pCO_2$—which washes nutrients into the oceans and causes eutrophication, leading to elevated oxygen consumption that depletes surrounding waters [Bond & Grasby (2017)]. High surface temperatures can also directly cause anoxic events by productivity stimulation [Bond & Grasby (2017)] and enhanced ocean stratification (which reduces ocean ventilation) [Erbacher et al. (2001)]. However, productivity stimulation from the direct injection of nutrients into the ocean by seafloor LIP eruptions can also cause anoxia [Ernst & Youbi (2017)], so a lack of $CO_2$-based warming does not necessarily preclude anoxic conditions from occurring. Similarly, thermogenic methane release and consequent rapid warming can be caused by an initial pulse of $CO_2$ warming, but it can also be caused by the intrusion of LIP magmas into organic-rich sediments and coal beds [Bond & Grasby (2017)]. A final example of this type is the depletion of the ozone layer, which leads to elevated surface ultraviolet fluxes and is suggested to have contributed to several mass extinctions via UV's ability to induce harmful elevated mutation rates [Beerling et al. (2007), Bond & Grasby (2017), Marshall et al. (2020)]. [Marshall et al. (2020)] suggest that rapid warming may induce ozone loss via enhanced convective transport of ClO

Corresponding Author: R.J. Graham, robert.graham@physics.ox.ac.uk

into the stratosphere, but other mechanisms have also been proposed to destroy the ozone layer, such as supernova-accelerated cosmic ray bombardment [Fields et al. (2020)], organohalogen release by LIP volcanism [Beerling et al. (2007), Broadley et al. (2018)], or proton bombardment from flaring by an active M-dwarf star [Segura et al. (2010), Tilley et al. (2019)].

Along with increased resilience to hyperthermal-related disasters, these results suggest that planets in the mid-to-outer reaches of the HZ have reduced susceptibility to global glaciation from transient increases in weathering. Changes to lithology, continent location, mountain-building, and many other processes can all cause weathering to accelerate or decline (e.g. [Kump et al. (2000), Penman et al. (2020)]), and workers have suggested that several partial [Macdonald et al. (2019)] and global [Goddéris et al. (2003), Donnadieu et al. (2004), Cox et al. (2016)] glaciations happened as a result of weathering acceleration. Just like $CO_2$ injection leads to smaller temperature increases under high background $pCO_2$, $CO_2$ sequestration leads to smaller temperature reductions under the same conditions, meaning this route to glaciation may be much less likely for high $pCO_2$ planets. Still, if excess weathering can be maintained on long enough timescales, ``limit cycling'' (the alternation between a warm climate state and a snowball climate state due to lack of stable weathering equilibrium [Mills et al. (2011), Kadoya & Tajika (2014), Menou (2015), Abbot (2016), Haqq-Misra et al. (2016)]) remains a possibility for these planets, since the $CO_2$ drawdown events from which mid-outer HZ planets are safe are transient pulses, not the quasi-permanent shifts to new carbon cycle equilibria required for limit cycle initiation. In addition, other non-weathering-related mechanisms like a rapid reduction in the level of a non-$CO_2$ greenhouse gas like methane [Kopp et al. (2005)] or a rapid increase in albedo due to stratospheric aerosol injection [Macdonald & Wordsworth (2017)] have also been proposed to trigger glaciation [Arnscheidt & Rothman (2020)]. However, the high Rayleigh scattering albedo (see Fig. 2) and consequent reduced sensitivity of planetary albedo to surface albedo in high-$pCO_2$ atmospheres means the ice-albedo feedback is less effective on these planets, since the growth of sea ice has less effect on the radiative balance of the planet under high $pCO_2$ [Pierrehumbert (2010), Von Paris et al. (2013)]. A set of calculations (not shown) that included a

Corresponding Author: R.J. Graham, robert.graham@physics.ox.ac.uk

temperature-dependent surface albedo feedback using the formula in [Wang & Stone (1980)] yielded qualitatively identical behavior to the constant-surface-albedo calculations presented in this paper for both $CO_2$ increases and $CO_2$ decreases, confirming the basic result that climate stability increases with decreasing instellation.

There are also a variety of catastrophe-inducing mechanisms that are not triggered by carbon cycle perturbations or their downstream effects. The frequencies of these events should not be affected by the presence of high background $pCO_2$. For example, extraterrestrial disaster sources like nearby supernovae (which may have a variety of negative impacts on atmospheric chemistry, e.g. [Reid et al. (1978)]), large asteroid/comet impacts [Alvarez et al. (1980), Hull et al. (2020)], and stellar superflares [Lingam & Loeb (2017)] likely do not discriminate by background $pCO_2$ levels. Similarly, global heavy metal poisoning by LIP volcanism [Lindström et al. (2019)] should remain a possible killing mechanism regardless of background $pCO_2$, though it is an interesting question whether this process alone could trigger a mass extinction without the additional environmental stressors from an accompanying hyperthermal episode. A variety of catastrophe triggers remain in play for mid-outer HZ planets despite the results of this study, but the main mass extinction mechanisms that have operated in Earth's past may be suppressed on these planets.

A corollary to the reasoning that planets at lower luminosity than the Earth are more stable to $CO_2$ perturbation-induced extinctions is that planets in the HZ receiving more instellation than Earth will be more susceptible to these perturbations because of their lower initial $pCO_2$ (although it is worth noting that Earth's position near the inner edge of the HZ means that there is not much space for temperate planets to reside at higher instellations). This point is noted in a blog post by Alex Tolley, who suggests that the increased volatility might be a serious enough barrier to habitability to justify revising the inner edge of the HZ outward to a slightly greater distance from the star [Tolley (2018)]. The same basic idea is elaborated by [Archer (2016)], who shows that anthropogenic climate

Corresponding Author: R.J. Graham, robert.graham@physics.ox.ac.uk

change would be progressing even more rapidly if Earth had possessed a lower initial $pCO_2$ when industrialization began and less rapidly if the initial $pCO_2$ had been higher.

In line with the logic of [Archer (2016)] and analogous to the reasoning presented in the discussion of previous mass extinctions on Earth, the results in this paper also show that civilizations arising throughout a large proportion of the HZ would be placed in no danger of self-induced climate change by burning fossil fuels. Since 5000 GtC is an approximate upper-bound on the accessible fossil fuel inventory on Earth [Ramirez et al. (2014)], civilizations arising anywhere beyond the inner boundary of the 5000 GtC stability zone will not face the degree of $CO_2$-based danger we face. Civilizational die-off from industrialization and pollution [Frank et al. (2018)] may be less of a danger on other worlds than it appears when extrapolating from Earth.

### 4.2. Does high $pCO_2$ preclude complex life?

A few studies have argued that high levels of atmospheric $CO_2$ may prevent the evolution of complex life on planets in the mid-outer reaches of the HZ [Bounama & Franck (2007), Schwieterman et al. (2019), Ramirez (2020)]. With the assumption of some threshold $CO_2$ beyond which complex life is not able to evolve, this idea can be used to define a ``habitable zone for complex life'' (HZCL), where temperatures can be maintained above freezing under the proposed $CO_2$ constraint. These arguments are based on observed responses of some forms of multicellular Earth-life, usually mammals, to high $CO_2$, low pH, and/or high CO (which can build up to high levels in high-$CO_2$ atmospheres of planets orbiting cool M or K stars [Schwieterman et al. (2019)]). It is important to note that the idea that CO might limit complex life is falsified by the fact that some extant invertebrates are invulnerable to CO's toxic effects because of their use of blood oxygen transport mechanisms that do not interact with CO [Howell (2019)]. Similarly, it is plausible that the pH- and $pCO_2$-sensitivity of the organisms that have been studied are not results of fundamental constraints on complex life, but are instead results of the specific chemical machinery employed by particular

Corresponding Author: R.J. Graham, robert.graham@physics.ox.ac.uk

Earth lineages, a point which is acknowledged in the studies that explore the HZCL idea [Schwieterman et al. (2019), Ramirez (2020)].

Still, if complex life is limited by high $pCO_2$, this has interesting implications for the results in this study. Using the most conservative value for maximum $pCO_2$ that appears in [Bounama & Franck (2007), Schwieterman et al. (2019), Ramirez (2020)], 0.01 bar, produces an outer edge for the HZCL of $S_{eff}$=0.76 with the climate model used in this paper, implying an outer radius of d = 1.15 au. This means there is no overlap between this conservative choice of HZCL and the 50,000 GtC stability zone defined in this paper (see Section 3.1), and only a narrow ring of overlap with the 5000 GtC stability zone (1.15-1.12 = 0.03 au). So, if the tolerance of complex life for $CO_2$ is low, this would suggest that much of the complex life in the universe could face carbon cycle volatility comparable to what the Earth experienced during the Phanerozoic Eon, conditional on significant similarities to Earth in terms of tectonics, mantle processes, and surface processes.

Alternatively, if we take the value for the outer edge of the HZCL to be at the luminosity where $pCO_2$=0.1 bar and $T$=273.15 K, perhaps based on the argument from lipid solubility theory in [Ramirez (2020)], the model used in this paper produces $S_{eff}$=0.67 and d = 1.22 au. With a 1D latitudinally-dependent energy balance climate model, [Ramirez (2020)] reports a slightly more distant boundary of d = 1.27 au with 1 bar background $N_2$. Our preference is for the latter value, since it comes from a more sophisticated model, in which case the HZCL encompasses the entire 5000 GtC stability zone, as well as a small ring with a width of 0.07 au in the interior portion for the 50,000 GtC stability zone. Higher choices of the toxic $pCO_2$ level lead to greater overlap with the 50,000 GtC stability zone. So, under the less conservative choices for maximum $pCO_2$, there is an optimal zone that allows for $CO_2$ levels that are high enough to reduce climate sensitivity to carbon cycle perturbations while being small enough to allow complex life to flourish. Planets in this zone might be considered ``superhabitable'' [Heller & Armstrong (2014)].

### 4.3. Implications for distribution of biosignatures?

Corresponding Author: R.J. Graham, robert.graham@physics.ox.ac.uk

As mentioned in the introduction, the largest known extinction in Earth's history is the Permian-Triassic event, a crisis apparently triggered by the eruption of a LIP, which eliminated up to 96% of marine species (and a significant but potentially smaller fraction of species on land) [Bond & Wignall (2014), Benton (2018)]. Several other mass extinctions that eliminated most of the species on Earth at the time also coincided with LIP eruptions [Kidder & Worsley (2010), Bond & Grasby (2017)]. The severity and frequency of these mass extinctions in Earth history raises the question of whether they present a danger to the persistence of complex biospheres over geologic timescales – it is conceivable that the Permian-Triassic extinction might have wiped out 100% of macroscopic species had the event been somewhat more severe.

If large, productive oxygenic biospheres are difficult to maintain for long periods because of extinction events, planets with detectable $O_2/O_3$ biosignatures may be few and far between, and thus difficult to find with near-future telescope surveys (see e.g. [Kawashima & Rugheimer (2019), Checlair et al. (2020)] for discussions of next-generation $O_2/O_3$ biosignature detection). In a work addressing the question of the persistence of Earth-like biospheres in the face of extinction events, [Tsumura (2020)] models Earth's extinction history as a random multiplicative process, and his methodology leads him to conclude that the probability of life's survival on Earth from the time of its origin to the present day is only 15% (since it is likely extremely difficult to extinguish prokaryotic life, this result should appropriately be considered to apply only to complex, multicellular life, not life as a whole). [Tsumura (2020)] suggests that this value can be used in models attempting to estimate the prevalence of intelligent life in the universe.

The results in this article demonstrate that Tsumura's calculations cannot be applied to infer the rate of complex life sterilization on Earth-like planets in general, even if they do accurately represent Earth itself. This is because the mechanisms that produced Earth's particular extinction history are strongly dependent on background atmospheric state, which is expected to systematically vary. In particular, the increased resilience to carbon cycle perturbations of planets in the mid-outer HZ

Corresponding Author: R.J. Graham, robert.graham@physics.ox.ac.uk

implies that these planets will be much less susceptible to the primary form of mass extinction that has operated on Earth, suggesting a higher long-term survival fraction for complex biospheres that occupy the stability zones outlined in Section 3.

Overall, this discussion implies that $O_2$/$O_3$ biosignatures may be more common in the middle-outer reaches of the HZ compared to the innermost region due to reduced likelihood of extinction because of greater environmental stability under high-p$CO_2$ atmospheres. For planets orbiting F-, G- and early-K-type stars that display significant brightening on gigayear timescales, this might suggest an anti-correlation between stellar age and biosignature prevalence because the increase in luminosity pushes the location of stability zones farther and farther away from the star over time, moving a larger and larger fraction of planets into the less-stable, more-extinction-prone interior region of the HZ. This prediction is directly opposite that of [Bixel & Apai (2020)], who note that the appearance of detectable $O_2$/$O_3$ biosignatures on Earth occurred one to two billion years after the origin of life, suggesting that planets orbiting older stars may be better candidates for biosignature searches.

An identical point regarding safety from catastrophic sterilization can be made about the distribution of technosignatures (biosignatures that imply the presence of technological civilizations e.g. [Lin et al. (2014)]). As noted in Section 4.1, analogues of anthropogenic climate change will be much less severe on planets in the stability zones defined in this paper, which suggests that civilizations on these planets may experience less stress from industrialization than those at higher instellations, perhaps allowing them to persist longer and increasing the likelihood of observable technosignatures in the stability zones.

Another wrinkle in this discussion is the possibility that non-sterilizing mass extinctions and climatic perturbations facilitate the emergence of novelty in the biosphere by freeing up or creating niches and restructuring ecosystems, perhaps allowing for greater complexity/biodiversity to be attained post-recovery e.g. [Wagner et al. (2006)]. An extreme example of this is the suggestion that

Corresponding Author: R.J. Graham, robert.graham@physics.ox.ac.uk

the Neoproterozoic snowball Earth episodes triggered the emergence of animal multicellularity [Simpson (2019)]. If climate perturbations do have a long-term positive effect on global biodiversity and novelty, then it is possible that planets in the stability zones will be less likely to have detectable biosignatures, depending on whether this effect is more important than the possibility of sterilization of complex life. However, the ecological and biogeochemical dynamics of post-catastrophe recovery are complex and poorly understood [Solé et al. (2002), Hull (2015)], and the paleontological data on the long-term impact of non-sterilizing catastrophes on global biodiversity is open to various interpretations [Michael Foote, personal communication].

So, biosignatures may be more likely in the high $pCO_2$ stability zones. This selection criterion can be used in conjunction with others, e.g. the HZCL [Schwieterman et al. (2019), Ramirez (2020)], to rank and prioritize targets in the HZ for next-generation telescopes to probe in search of life. If decreased likelihood of catastrophe leads to a greater likelihood of detectable biosignatures, a large enough sample of $O_2/O_3$ detections might reveal a trend with a higher-than-expected fraction of biosignature detections at low instellations within the HZ.

### 4.4. Implications for Earth history?

Although there is still considerable debate about Earth's $CO_2$ levels throughout the past 4 Ga (e.g. [Charnay et al. (2020)]), recent proxy constraints suggest that $CO_2$ partial pressure may have been approximately 1000x its present value (perhaps ~0.1-0.5 bar) in the late Archean [Lehmer et al. (2020), Payne et al. (2020)], with a gradual decline over the Proterozoic Eon driven by the response of weathering to a brightening Sun [Kanzaki & Murakami (2015), Krissansen-Totton et al. (2018), Krissansen-Totton & Catling (2020)]. The role of other greenhouse gases during the Precambrian is unclear, but there is some evidence for strongly elevated methane levels during the Archean, which may also have accounted for some of the warming needed to offset the climate impacts of the dimmer young Sun [Catling & Zahnle (2020)]. Probabilistic models of ocean pH through geologic time that account for the wide uncertainties in relevant proxy data produce a corresponding history with

Corresponding Author: R.J. Graham, robert.graham@physics.ox.ac.uk

pH increasing from an initial slightly acidic-to-neutral range of 6.5-7 in the Archean to a slightly basic range of 7.5 to 9 during the Phanerozoic [Halevy & Bachan (2017), Krissansen-Totton et al. (2018), Krissansen-Totton & Catling (2020)].

High values of $pCO_2$ and low values of ocean pH in Earth's past would suggest that the planet may have exhibited the stability to $CO_2$ perturbations described in this paper during the Archean and part of the Proterozoic. Much of the Precambrian seems to have displayed notable climatic stability and warmth, particularly after the major period of global glaciation leading from the Archean into the Proterozoic, which may have marked the transition from an oxygen-poor atmosphere with strong greenhouse warming from both $CO_2$ and $CH_4$ to an oxidized atmosphere with limited $CH_4$ warming [Daines & Lenton (2016), Olson et al. (2016), Catling & Zahnle (2020)]. With $pCO_2$ held at high levels (perhaps 0.01-0.1 bar) by the silicate weathering feedback [Krissansen-Totton et al. (2018)] and, potentially, $CO_2$-generating authigenic clay formation (``reverse weathering'') [Isson & Planavsky (2018), Krissansen-Totton & Catling (2020)], the mid-Proterozoic atmosphere would have been quite stable to transient $CO_2$ release and/or consumption from the many LIPs recorded during this period [Ernst & Youbi (2017)]. This may partially explain the apparent lack of transient climate extremes during the ``Boring Billion'' period [Buick et al. (1995)] between approximately 1.7 to 0.75 Ga, though underlying geologic causes are still necessary to explain the long-lasting warm, steady background climate [Cawood & Hawkesworth (2014)]. The stability to $CO_2$ perturbations would have decreased over time as increasing solar luminosity caused $CO_2$ to be drawn down, which may have been a necessary precondition that allowed for transient enhanced weathering events to trigger one or both of the Neoproterozoic snowball episodes [Goddéris et al. (2003), Donnadieu et al. (2004)]. Even in the Phanerozoic, the issue of background $pCO_2$ and its influence on the sensitivity of climate to a given increase/decrease in $CO_2$ plays an important role in the debate over the interpretation of the evidence for events like the Paleocene-Eocene Thermal Maximum [Pagani et al. (2006), Higgins & Schrag (2006)].

Corresponding Author: R.J. Graham, robert.graham@physics.ox.ac.uk

The remaining lifespan of the complex biosphere is estimated to be ~1 Ga based on the modeled timing of the drawdown of atmospheric $CO_2$ to a level below the lower limits of tolerance for C4 plants as the Sun continues to brighten [Caldeira & Kasting (1992)]. However, a severe LIP eruption could conceivably end complex life much earlier, similar to the suggestion that a LIP eruption ended habitable conditions on Venus [Ernst et al. (2017), Way & Del Genio (2020)]. The low $CO_2$ conditions of Earth's future should make environmental conditions even more sensitive to $CO_2$ perturbations than they were during past LIPs. However, the evolution of pelagic calcifiers over the past 200 million years may have strongly increased environmental resilience to $CO_2$ perturbations [Henehan et al. (2016)], suggesting a potential counter to the effects of reduced $CO_2$. A biogeochemical model that includes the effects of pelagic calcifiers on ocean chemistry could be used to examine the impact of future LIP eruptions under low $pCO_2$, high insolation conditions.

## 5. Summary & Conclusions

In this article, we used idealized climate and ocean chemistry models to demonstrate that high $pCO_2$ atmospheres on planets in the middle and outer regions of the HZ confer robust temperature- and pH-stability in the face of $CO_2$ perturbations that would cause severe environmental change in the low-$pCO_2$ context of Phanerozoic Earth. Both temperature- and pH-stability arise because of the rapid growth in $pCO_2$ necessary to maintain a temperate climate as instellation decreases with distance from the host star on planets with $N_2$-$CO_2$-$H_2O$ atmospheres in the HZ. The $pCO_2$ necessary for above-freezing temperatures at the outer edge of the HZ can be more than a million times greater than that required near the inner edge, assuming constant $pN_2$, so $CO_2$'s logarithmic radiative forcing response means that a given mass of $CO_2$ added to or subtracted from the atmosphere of a temperate planet near the inner edge has a vastly larger effect on surface temperature than the same change in atmospheric $CO_2$ mass near the outer edge. Similarly, as $pCO_2$ increases, there is a reduction in the fraction of $CO_2$ that partitions into the ocean as dissolved inorganic carbon for a given

Corresponding Author: R.J. Graham, robert.graham@physics.ox.ac.uk

mass of carbon injected into the atmosphere-ocean system, meaning the marginal impact of a given $CO_2$ change on ocean pH decreases as background levels go up.

The $CO_2$ perturbations imposed in the experiments in this study are equivalent to 5000 GtC and 50,000 GtC added to the atmosphere, which are, respectively, an estimate of the amount of carbon that would be released if all accessible fossil fuels were burnt [Ramirez et al. (2014)] and a high estimate of the amount of carbon released due to LIP volcanism during the most severe mass extinction in Earth history [Kump (2018)]. The temperature response to the 5000 GtC perturbation falls below 4 K at $S_{eff}=0.8$, where the initial $pCO_2=1.5 \times 10^{-3}$ bar, corresponding to an orbital distance of $d = 1.12$ au in the Solar System. The temperature response to the 50,000 GtC perturbation reaches the same 4 K threshold at $S_{eff}=0.7$ and $pCO_2=5.4\times10^{-2}$ bar, corresponding to $d=1.2$ au. Those values of $d$ can serve as inner boundaries for ``stability zones'' where $N_2$-$CO_2$-$H_2O$ atmospheres are relatively invulnerable to temperature effects of $CO_2$ changes at or below a given magnitude (see Fig. 5). Analogously, the magnitude of pH reduction falls below 0.5 units at $S_{eff}=0.82$ ($d=1.1$ au) and $S_{eff}=0.78$ ($d=1.13$ au) for 5000 GtC and 50,000 GtC perturbations respectively.

These results suggest that many Earth-like planets throughout the mid-to-outer habitable zone may be safe from extreme environmental change from transient $CO_2$ perturbations of sizes that have caused catastrophes in Earth's past, particularly planets in the 50,000 GtC stability zone (Figs. 4, 5). Similarly, climate change and ocean acidification from $CO_2$ emitted by the burning of fossil fuels may be a less serious issue for civilizations on planets receiving less instellation than Earth (see 5000 GtC stability zone in Fig. 5). This represents a significant habitability advantage for temperate planets in the mid-to-outer reaches of the HZ compared to those near the inner edge. The large climate swings that have punctuated the Phanerozoic Eon are in part a result of Earth's low $pCO_2$ and relative proximity to the inner edge of the habitable zone during this period.

## Acknowledgments
I thank Ray Pierrehumbert, Dorian Abbot, Michael Foote, and Sarah Rugheimer for useful discussions over the

Corresponding Author: R.J. Graham, robert.graham@physics.ox.ac.uk

course of the writing of this paper. I also thank Ray, Dorian, Sarah, and two anonymous reviewers for helpful comments on various versions of the manuscript.

Disclosure statement

I have no conflicts of interest.

Funding

I gratefully acknowledge scholarship funding from the Clarendon Fund and Jesus College, Oxford.

Corresponding Author: R.J. Graham, robert.graham@physics.ox.ac.uk

Corresponding Author: R.J. Graham, robert.graham@physics.ox.ac.uk

Corresponding Author: R.J. Graham, robert.graham@physics.ox.ac.uk

Corresponding Author: R.J. Graham, robert.graham@physics.ox.ac.uk

Corresponding Author: R.J. Graham, robert.graham@physics.ox.ac.uk

Corresponding Author: R.J. Graham, robert.graham@physics.ox.ac.uk

Corresponding Author: R.J. Graham, robert.graham@physics.ox.ac.uk

Corresponding Author: R.J. Graham, robert.graham@physics.ox.ac.uk